\documentstyle[aps,prb,multicol,epsf,rotate]{revtex}

\def\vc{{\bf c}}
\def\vd{{\bf d}}
\def\vf{{\bf f}}
\def\vp{{\bf p}}
\def\vq{{\bf q}}
\def\vv{{\bf v}}

\def\vz{{\bf z}}

\def\vH{{\bf H}}

\newcommand{\vvarepsilon}{\mbox{\boldmath$\varepsilon$}}
\newcommand{\vDelta}{\mbox{\boldmath$\Delta$}}
\newcommand{\vsigma}{\mbox{\boldmath$\sigma$}}

\draft

\begin{document}

\title{Identifying the pairing symmetry in the Sr$_2$RuO$_4$ superconductor}
\author{Matthias J. Graf and A.~V. Balatsky}
\address{Theoretical Division, Los Alamos National Laboratory, 
Los Alamos, New Mexico 87545}

\date{30 May 2000}
\maketitle

\begin{abstract}
We have analyzed heat capacity and thermal conductivity measurements of
Sr$_2$RuO$_4$ in the normal and superconducting state 
and come to the conclusion
that an order parameter with nodal lines on the Fermi surface is required
to account for the observed low-temperature behavior. A gapped order parameter
is inconsistent with the reported thermodynamic and transport data.
Guided by a strongly peaked dynamical susceptibility along the diagonals of 
the Brillouin zone in neutron-scattering data, we suggest a spin-fluctuation
mechanism that would favor the pairing state with the gap maxima along the 
zone diagonals (such as for a $d_{xy}$ gap). The most plausible candidates 
are an odd parity, spin-triplet, $f$-wave pairing state, 
or an even parity, spin-singlet, $d$-wave state.
Based on our analysis of possible pairing functions we propose measurements
of the ultrasound attenuation and thermal conductivity in the magnetic field 
to further constrain the list of possible pairing states.
\end{abstract}

\pacs{PACS numbers: 74.25.Fy, 74.25.Bt, 74.25.Ld \hfill LA-UR:00-1398}

\begin{multicols}{2}

\section{Introduction}

The search for the superconducting pairing symmetry
in the layered perovskite material Sr$_2$RuO$_4$ (SrRuO),
and its attempted theoretical predictions, 
show remarkable parallels to the heavy-fermion superconductor UPt$_3$. 
In both systems, early specific-heat measurements 
showed a large residual value of $C/T$ at low temperatures and were 
interpreted in terms of a superconducting phase with a nonunitary 
$p$-wave order 
parameter.\cite{specheat,nishizaki97,machida89,machida96,sigrist96} 
The observation of a strong $T_c$ suppression with nonmagnetic 
impurities was an additional indication of a superconducting phase
with an unconventional order 
parameter.\cite{vorenkamp93,dalichaouch95,kycia98,mackenzie98}
However, newer measurements on high-quality single crystals have 
shown that the most likely pairing state in UPt$_3$ is an 
$f$-wave state, or more precisely a spin-triplet state whose orbital
basis function belongs to the $E_{2u}$ representation of the hexagonal
crystallographic point group (D$_{6h}$).\cite{machida99b,graf00}
The experience with UPt$_3$ suggests that the early identification of the 
pairing state, based on low-quality, inhomogeneous samples,
is at best inconclusive (for a review on UPt$_3$ see, for example, 
Refs.~\onlinecite{heffner96,sauls94}). 
However, with improving sample quality it  becomes feasible to 
identify the pairing state by studying transport properties.

 Here we analyze new heat capacity measurements on high-quality single crystals
of SrRuO, as well as thermal conductivity data on dirty 
samples with a strong $T_c$ suppression, to show that the proposed 
$p$-wave model,\cite{rice95,agterberg97,sigrist99}
$\vDelta(\vp_f) \sim (p_x + i p_y) \hat{\vz}$,
is inconsistent with the 
available data. Our conclusion is that the pairing state  in SrRuO,
most likely, has lines of nodes with gap nodes given by the
$d_{xy}$ gap function. 
This can occur in either  an $f$-wave state, i.e., a spin-triplet
pairing state belonging to the $E_u$ representation of the tetragonal
crystallographic point group (D$_{4h}$) or in a $d_{xy}$ singlet state. 
We argue that the $f$-wave nodal state is consistent with measurements
of the heat capacity,\cite{nishizaki97}
thermal conductivity,\cite{suderow98,tanatar00}
penetration depth,\cite{bonalde00} 
Andreev reflection,\cite{laube00}
NMR,\cite{ishida97}
Knight shift,\cite{ishida98}
and $\mu$SR experiments.\cite{luke98}

\begin{figure}
\noindent
\begin{minipage}{86mm}
\epsfxsize=60mm
\centerline{ {\epsfbox{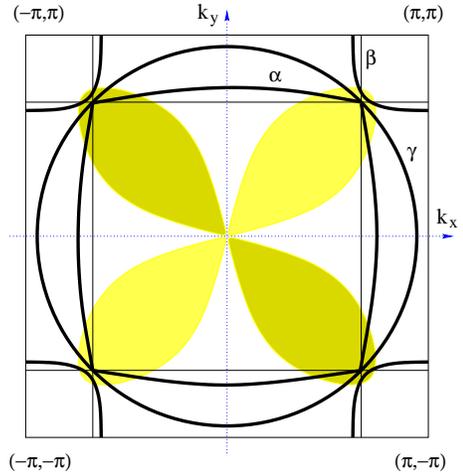}} }
\vspace*{5pt}
\caption[]{Fermi surfaces in the Brillouin zone after Mazin and
Singh.\cite{mazin97}  The plotted order parameter
(proportional to $d_{xy}$) opens a gap along $(\pm \pi, \pm \pi)$
where the incommensurate peaks of the spin susceptibility are observed.
The corresponding quasi-one-dimensional model bands $\xi$ at $(k_x, \pm 2\pi/3)$
and $\zeta$ at $(\pm 2\pi/3, k_y)$ are shown as thin lines.
}\label{BZ}
\end{minipage}
\end{figure}

Recent band-structure calculations by Mazin and Singh\cite{mazin97} 
indicate that 
there is an increase in the spin susceptibility $\chi({\bf q},\omega)$ 
at four points in the Brillouin zone at approximately 
$\vq_0 \approx (\pm 2\pi/3, \pm 2\pi/3)$ that occur due to strong nesting
effects of quasi-one-dimensional bands ($\xi$ and $\zeta$).
Nesting effects among these bands lead to the increased interaction 
between particles on the Fermi surface near $\vq_0$, see Fig.~\ref{BZ}.
In recent neutron-scattering experiments\cite{sidis99} the predicted 
four incommensurate peaks near $\vq_0$ were indeed observed thus supporting 
that nesting effects near these points are 
important.\cite{mazin97}

In this paper we propose: 
1)  To identify the regions at the Fermi surface near $\vq_0$ 
with the ones that develop the largest gap. 
We use the neutron-scattering data as an indication that near the nesting 
regions the particle-particle (or particle-hole)
interactions are dominant and that these are 
the regions that would benefit the most from opening a
superconducting gap. 
2) We suggest that regardless of the singlet or triplet nature of the pairing 
in SrRuO the gap function should  be proportional to a $d_{xy}$ harmonics.
Such an order parameter would lead to {\em lines of nodes} along the
$k_z$-axis in the gap and to power-law behavior in the thermodynamic and 
transport properties.
Line nodes on the Fermi surface lead in clean superconductors, and for 
scattering in the Born limit, to the well-known temperature dependences
\cite{pethick86,schmitt_rink86,hirschfeld86,monien87}
of the specific heat 
$C \sim T^2$, the nuclear spin relaxation rate $1/T_1 \sim T^3$, 
the deviation of the penetration depth from its zero-temperature value
$\triangle\lambda \sim T$, 
the thermal conductivity $\kappa \sim T$, 
and the longitudinal sound attenuation 
$\alpha_L \sim {\rm const}.$, as well as for the transverse 
attenuation $\alpha_T \sim T^2$.
3) Based on the proposed line nodes in the gap we make predictions 
for ultrasound attenuation and thermal conductivity measurements
that can further distinguish between the remaining possible basis
functions. 
We propose complimentary longitudinal and transverse attenuation measurements 
that can help to locate the location of the nodal lines of the order parameter 
on the Fermi surface.
Another crucial experiment is the thermal conductivity with an 
in-plane magnetic field.  We expect the {\em  fourfold} modulation of the
thermal conductivity $\kappa(\theta, H)$ as a function of the angle
between the nodes of the gap [along the (1,0) and (0,1) direction] and the
field directions.  Thermal conductivity measures the
unpaired quasiparticle heat transport and is therefore sensitive to
the angular (field) dependence of the quasiparticle scattering 
rate, which ``{knows}'' about the angular dependence of the gap.
We use the analogy with the suggested d-wave paring state in high-$T_c$
superconductors where this {\em  fourfold} modulation has been
observed.\cite{yu95,aubin97,hirschfeld98,vekhter99}

\section{Model}

The gap function for even parity (spin-singlet) or 
odd parity (spin-triplet) representations
is described by an order parameter of the form
\begin{eqnarray}
&\Delta_{\alpha \beta}({\bf p}_f)  =  \Delta(\vp_f)
(i \sigma_y)_{\alpha \beta} \,,\quad & {\rm (singlet)}
\\
&\Delta_{\alpha \beta}({\bf p}_f)  =  \vDelta(\vp_f) \cdot 
(i \vsigma \sigma_y)_{\alpha \beta} \,,\quad & {\rm (triplet)}
\end{eqnarray}
with $\sigma_\alpha$ being Pauli matrices. Since nonunitary
states,\cite{sigrist96,machida96}
i.e., $\vDelta \times \vDelta^* \ne 0$, have been ruled
out by the very small residual value of the specific heat $C/T$ at zero
temperature,\cite{nishizaki99} we restrict our study of spin-triplet states 
to unitary order parameters that factorize into a single spin vector
and an orbital amplitude, i.e., $\vDelta(\vp_f) = \vd \,\Delta(\vp_f)$,
where $\vd$ is a real unit vector in spin space and $\Delta(\vp_f)$
is an odd-parity orbital function. The vector $\vd$ defines the axis
along which the Cooper pairs have zero spin projection, e.g., if 
$\vd || \hat{\vz}$, then $\Delta_{\uparrow \uparrow} = 
\Delta_{\downarrow \downarrow} = 0$ and $\Delta_{\uparrow \downarrow} =
\Delta_{\downarrow \uparrow} = \Delta(\vp_f)$.

Whether or not spin-orbit coupling is weak or strong in Sr$_2$RuO$_4$
has important ramifications for both spin and orbital components of
the order parameter that are allowed by symmetry. While spin-orbit
coupling is believed to be strong in the heavy-fermion system UPt$_3$
there are no experimental indications that this is likewise true
for Sr$_2$RuO$_4$. In the meantime we will use the classification
of basis functions in terms of irreducible representations of the 
tetragonal point group ($D_{4h}$) listed in Table~\ref{basis},
implying that spin-orbit coupling is strong. Since the band-structure
calculations\cite{mazin97} 
and de Haas - van Alphen measurements\cite{mackenzie98b}
show very little 
dispersion along $k_z$, we will consider only two-dimensional (2D)
basis functions on a more or less cylindrical Fermi surface.
A similar list of possible basis functions was recently
compiled by Hasegawa and co-workers\cite{hasegawa99}
for further investigations.
The listed {\it hybrid} state (\#3) of the direct product 
$B_{2g} \otimes E_u = E_u$ is a
non-trivial realization of the $E_u$ representation
(also referred to as $f$-wave state).
So far Knight shift data with an in-plane magnetic field
$\vH || [100]$ show no change below $T_c$ and have been interpreted 
in terms of spin-triplet pairing with the spin vector $\vd$ locked 
to the crystal $\vc$-axis.\cite{ishida98}
On the other hand, muon spin rotation ($\mu$SR)
experiments observed a spontaneous
internal magnetic field on entering the superconducting state,\cite{luke98} 
consistent with a time-reversal symmetry breaking state belonging 
to the two-dimensional $E_u$ representation. 

\noindent
\begin{minipage}{86mm}
\noindent
\begin{table}
\caption[]{2D polynomial basis functions for the irreducible representations
of $D_{4h}$ of several pairing models (after Yip and Garg\cite{yip93}). 
Notice that $B_{g} \times E_u = E_u$. 
The commonly proposed $p_x + i p_y$ state belongs to the two-dimensional 
$E_u$ representation. We present {\em both} singlet, $d_{xy}$, and triplet
states, $B_{2g} \otimes E_u$, which have lines of nodes, 
as plausible candidates for Sr$_2$RuO$_4$. For simplicity we list only the
nodal angles on the dominating $\gamma$ and $\alpha$ Fermi sheets.
}
\label{basis}
\begin{tabular}{cccc}
\# 	& $\Gamma$	& $\Delta(\vp_f)$	& nodes
\\ \hline
1	& $B_{2g}$	& $p_x p_y$ 		& 
$\phi = 0, \pi/2, \pi, 3\pi/2$
\\
2	& $E_u$		& $(p_x + i p_y) $	& no
\\
3	& $B_{2g} \otimes E_u$	& $p_x p_y (p_x + i p_y) $	&
$\phi = 0, \pi/2, \pi, 3\pi/2$
\end{tabular}
\end{table}
\end{minipage}

At this place a caveat is warranted because
neither Knight shift data at high fields and for a single field orientation,
nor $\mu$SR measurements in impure samples provide a clear-cut identification 
for spin-triplet pairing or broken time reversal symmetry states. 
For example, in UPt$_3$ early $\mu$SR measurements indicated broken 
time-reversal symmetry in the superconducting phase (probably due to
impurities), while newer measurements on very clean samples fail to 
detect any effect at all.\cite{yaouanc00}
What makes the interpretation of the Knight shift data in SrRuO
for magnetic fields parallel to the planes even more complicated,
is, that (1) the experiment was not performed
in the low field limit, but rather deep in the mixed phase, 
$H \sim H_{c2}/2$, where contributions from the vortices may be important, 
and (2) nonlocal and surface effects may be relevant due to a small 
Ginzburg-Landau parameter for in-plane currents,
$\kappa_{||} = \lambda_{||}/\xi_{||} \sim 2.6$.

\section{Thermodynamic and transport properties}

We calculate the specific heat and thermal conductivity for the 
order-parameter models listed in Table~\ref{basis} and fit the results to existing 
experiments. This way, we can determine the model parameters and make
predictions for sound attenuation measurements.
It is important to point out that none of the here
analyzed transport experiments can distinguish between a spin-singlet 
and a spin triplet order parameter. Thus we obtain identical results
for the states \#1 and \#3.

The specific heat, $C = T dS/dT$, can easily be obtained from the 
entropy,\cite{rickayzen69,monien87}
\begin{equation}
S = 4 \int_0^\infty d\epsilon N(\epsilon)
\left(
 \frac{\epsilon}{T} f(\epsilon) - \ln [1-f(\epsilon)]
\right)
\,,
\end{equation}
by numerical differentiation. Here 
$f(\epsilon)=1/[1+\exp(\epsilon/ T)]$ is the Fermi-Dirac function
and $N(\epsilon) = - (N_f/\pi) {\rm Im}\int d\vp_f g^R(\vp_f,\epsilon)$ is
the density of states per spin with $N_f$ being the normal-state density 
of states at the Fermi surface. 

In the limit of Born (weak) or unitarity (strong) impurity scattering
the in-plane thermal conductivity of unitary spin-triplet superconductors
is given by\cite{graf96,graf}
\begin{eqnarray}
\kappa_{i i} &=& -\frac{N_f v_f^2}{8\pi^3 T^2}
\int\! d\epsilon \, \epsilon^2 \, {\rm sech}^2 \frac{\epsilon}{2 T}
\int\! d\vp_f {\hat{\vv}_{fi}^2} {\cal K}(\vp_f, \epsilon)
\,,
\end{eqnarray}

\begin{eqnarray}
{\cal K}(\vp_f, \epsilon) = 
\frac{1}{{\rm Re}\, C^R} 
\Big[ g^R (g^R)^* - \vf^R \cdot (\vf^R)^* + \pi^2 \Big]
\,,
\end{eqnarray}
with the unit vector of the Fermi velocity, $\hat{\vv}_{fi}$, and
$C^R = -\frac{1}{\pi}\sqrt{ |\vDelta|^2 - (\tilde\epsilon^R)^2}$.
The quasiclassical equilibrium Green functions are 
$g^R = \tilde\epsilon^R/C^R$ and $\vf^R = -\vDelta/C^R$.
Within the $t$-matrix approximation for isotropic scattering
the impurity renormalized quasiparticle energy is 
$\tilde\epsilon^R = \epsilon - \sigma^R_{imp}(\epsilon)$.
For weak scattering 
$\sigma^R_{imp}(\epsilon) = (\Gamma/\pi) \int d\vp_f g^R$,
and for strong scattering 
$\sigma^R_{imp}(\epsilon) = -\Gamma /  (\pi \int d\vp_f g^R$),
with the normal-state scattering rate $\Gamma = \hbar/2\tau$.

In the hydrodynamic regime, $\omega \tau \ll 1$, and long wavelength limit,
$q \ell \ll 1$, the absorption of ultrasound of polarization
${\vvarepsilon}$ propagating along direction $\vq$ is related
to the viscosity by\cite{schmitt_rink86,hirschfeld86,monien87}
\begin{equation}
\alpha = \frac{\omega^2}{\varrho c_s^3} \eta_{ij,kl} 
\hat{\vvarepsilon}_i \hat{\vq}_j \hat{\vvarepsilon}_k \hat{\vq}_l
\,,
\end{equation}
with the speed of sound $c_s$, the mass density $\varrho$, and 
the viscosity tensor evaluated at $\omega \to 0$,
\begin{eqnarray}\label{viscosity}
\eta_{ij,kl} =
-\frac{N_f v_f^2 p_f^2}{8\pi^3 T}
\int\! d\epsilon \, {\rm sech}^2 \frac{\epsilon}{2 T}
\int\! d\vp_f \pi_{ij}\pi_{kl}
{\cal K}(\vp_f, \epsilon)
\,,
\nonumber \\
\end{eqnarray}
where
$\pi_{ij} = {\hat{\vv}}_{fi} {\hat{\vp}}_{fj} - {1\over 2}\delta_{ij}$.

Here we confine our discussion to order parameters with vanishing
averages, $\int d\vp_f\, \vDelta(\vp_f) = 0$, which satisfy the
gap equation for triplet (singlet) pairing interactions,
\begin{equation}
\vDelta(\vp_f) = \int \frac{d\epsilon}{2\pi} \tanh\frac{\epsilon}{2 T}
\int\! d\vp_f^\prime V(\vp_f,\vp_f^\prime) 
{\rm Im}\,\vf^R(\vp_f^\prime, \epsilon)
\,.
\end{equation}
Note that for spin-singlet pairing all vector functions get 
replaced by the corresponding scalar functions.
In the weak-coupling spin-fluctuation model the pairing interaction 
is written as
\begin{eqnarray}
& V(\vp_f,\vp_f^\prime) \sim 
V^*( \vp_f, \vp_f' ) \chi(\vp_f - \vp_f^\prime) \,, &\\
& \chi(\vq) = \chi_0 / \left[ 1 + \xi^2 ( \vq - \vq_0 )^2 \right] \,.&
\end{eqnarray}
The detailed form of the effective pairing interaction 
$V^*( \vp_f, \vp_f')$ 
depends on the form of the spin singlet or spin triplet pairing
interaction.
$\chi_0$ is the static spin susceptibility,
$\xi$ is the antiferromagnetic correlation length, and the incommensurate 
wave vectors are $\vq_0 \approx (\pm 2\pi/3, \pm 2\pi/3)$.
The spin-fluctuation scenario proposed here is similar to the 
one studied by many authors in the context of
the heavy-fermion systems,\cite{norman90}
the high-$T_c$ cuprates,\cite{spinfluctuations}
the quasi-two-dimensional organic superconductors,\cite{schmalian98}
and even SrRuO.\cite{mazin97,miyake99}
In contrast to the microscopic model calculations
in Refs.~\onlinecite{mazin97,miyake99},
we propose the existence of either an attractive
triplet $f$-wave or singlet $d$-wave pairing channel in order to 
describe the power-laws observed in thermodynamic and transport coefficients.
Our approach is guided by neutron-scattering data of the spin susceptibility 
that can lead to a gap function that is gapped along the 
$(\pm\pi,\pm\pi)$ directions and has nodes along $(\pi,0)$ and 
$(0,\pi)$.
The immediate consequence of the proposed state is that the
superconducting gap on the holelike $\beta$ band develops nodes at
$(\pm 2\pi/3, \pm \pi)$ and $(\pm \pi, \pm 2\pi/3)$.

\section{Results And Discussions}

In our analysis of the thermodynamic and transport properties we make
the simplifying assumption that all three Fermi surfaces 
$(\alpha, \beta, \gamma)$ simultaneously go superconducting 
and can be described by one  effective, cylindrical band.
At the present time we cannot rule out any admixture of the $p$-wave
state \#2 to the $f$-wave state \#3, since both gap functions belong
to the same two-dimensional representation $E_u$. 
However, from a detailed analysis of the calculated heat capacity 
we find, rather conservatively, that the admixture of a nodeless 
$p$-wave state has to be less than 20\% to be consistent with the 
experimental $C(T)$.\cite{graf_unpublished}
Thus, we neglect the possibility of a $p$-wave admixture to the 
$f$-wave gap function in the remainder of this work.
Impurity calculations for the $p$-wave state \#2  also
were performed by Maki and Puchkaryov,\cite{maki00a}
who reported reasonably good agreement between experiment and the
calculated $p$-wave order parameter.
Very recently, Dahm, Won, and Maki\cite{dahm00} discarded the nodeless 
$p$-wave state and argued in favor of $f$-wave pairing.

In the temperature range $T^* \ll T \ll T_c$, where $T^*$ is the
characteristic temperature of the impurity band width, and in the
clean limit, $\Gamma \ll \Delta_0$, the evaluation of the entropy and
transport coefficients simplifies significantly.
In the presence of line nodes on the Fermi surface the density of states is
$N(\epsilon) \sim (\epsilon/\Delta_0) N_f$. 
Similarly, we obtain for the Fermi surface averaged integrand
$\int d\vp_f \,{\cal K}(\vp_f, \epsilon) \sim 
 - \epsilon \tau(\epsilon)/\Delta_0 $,
because of
${\cal K} \approx \pi^4/({\rm Re}\,\tilde\epsilon^R {\rm Im}\,\sigma^R_{imp}) 
 {\rm Im}\, C^R$, with
$\tilde\epsilon^R \approx \epsilon + i 0^+$. 
Thus, the transport coefficients
show the usual power laws of clean superconductors when using
the approximate relations for the scattering self-energies,
$1/2\tau(\epsilon) = (-\pi)^{-1} {\rm Im}\,\sigma^R_{imp} 
 \sim \Gamma \epsilon/\Delta_0$ 
in the Born limit, or
$1/2\tau(\epsilon) \sim \Gamma \Delta_0/\epsilon$ 
for unitarity scattering.

\subsection{Specific heat}

\begin{figure}
\noindent
\begin{minipage}{86mm}
\epsfxsize=75mm
\centerline{ {\epsfbox{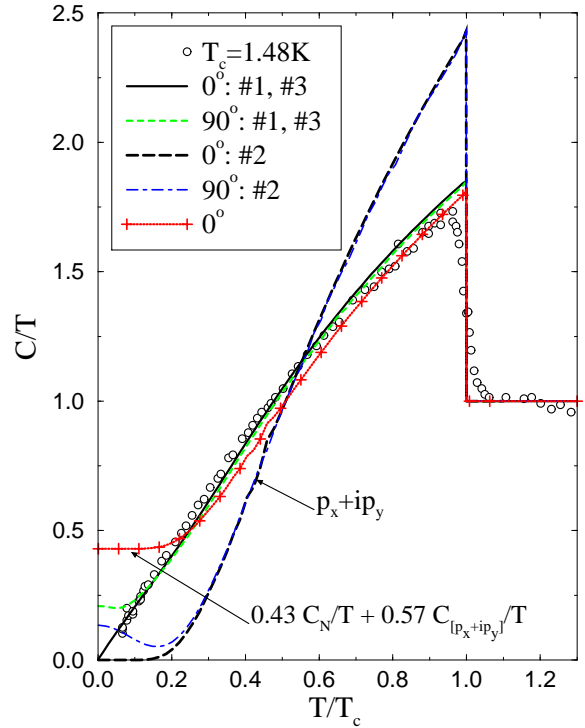}} }
\caption[]{The specific heat normalized at $T_c$
for pairing states with line nodes (\#1 or \#3).
A scattering phase shift of 
$\delta_0 = 0^o$ (Born) or $\delta_0 = 90^o$ (resonant), 
a scattering rate $\Gamma/\pi T_{c0} = 0.01$,
and a nodal parameter $\mu = 1.5$, were assumed.
For comparison the $p$-wave state \#2 for Born (long-dash)
and resonant (dot-dash) scattering and the multiband state
by Agterberg in the Born limit (cross-dot) are shown. 
The data are from Ref.~\onlinecite{nishizaki99}.
}\label{heat}
\end{minipage}
\end{figure}

The states \#1 and \#3 with line nodes yield 
$C \sim N_f T^2 / \mu \Delta_0$, 
in excellent agreement with experiments, while the gapped
state \#2 disagrees with the data.
The proposed multiband order-parameter model by Agterberg and
co-workers\cite{agterberg97}, which assumes that only one 
band $(\gamma)$ out of three possible bands goes superconducting at $T_c$,
fails to describe the low-$T$ dependence (see Fig.~\ref{heat}). 
Our result for the $p$-wave state \#2 is in agreement with calculations 
of the heat capacity by Agterberg.\cite{agterberg99}
In the multiband model the density of states (DOS) of the $\gamma$ band
is weighted with 57\% of the total DOS, while the remaining $\alpha$ and
$\beta$ bands account for 43\% of the total DOS. It is the $\gamma$ band
on which the $p$-wave state \#2 has been proposed to nucleate. The $\alpha$
and $\beta$ bands remain normal.
Here $\mu$ is the slope parameter of the gap function at the nodes, 
$\mu = |d\Delta(\phi)/\Delta_0 d\phi |_{node}$.
In our calculations we have used variational basis functions,
$\vDelta(\vp_f) \to \vDelta(\vp_f) {\cal F}_{A_{1g}}(\vp_f; \mu)$,
where the variational function ${\cal F}_{A_{1g}}$ belongs to the $A_{1g}$
representation and remains invariant under all group transformations. 
The slope parameter $\mu$ allows us to adjust the opening of the gap function
at the nodes, which is otherwise not determined by symmetry. This enables
us to quantitatively describe the ground state of the superconducting order
parameter as probed by low energetic quasiparticles.
An approach that has been quite successful in describing the low energetic
quasiparticle excitations in UPt$_3$.\cite{graf}

Assuming that pure SrRuO has an optimal transition temperature of 
$T_{c0} \simeq 1.51\, {\rm K}$,\cite{mackenzie98}
we obtain an excellent fit for scattering in
the Born limit with a scattering phase shift $\delta_0 \to 0$ and
a scattering rate $\Gamma / \pi T_{c0} = 0.01$.
On the other hand, resonant scattering $(\delta_0 \to \pi/2)$ with the same
scattering rate gives a residual value of $C/T$ that is too large.
If impurity scattering is indeed resonant,  then a value of
$\Gamma / \pi T_{c0} \le 10^{-3}$ is required to account for the lowest
measured values of the specific heat. Furthermore, 
it would imply that the optimal
transition temperature is closer to $T_{c0} \simeq 1.48 \, {\rm K}$.

The $T_c$ transition of the two components of
the triplet $p$-wave order parameter \#2,
or of the two components of the $f$-wave order parameter \#3, 
is doubly degenerate.
Similar to the multicomponent superconducting order parameter
in UPt$_3$ uniaxial strain (pressure) in the plane would lift the 
degeneracy of the two-component order parameter.\cite{jin97}
As a consequence the transition temperature will split into two. 
This is a crucial test of the multicomponent nature of the
order parameter.\cite{sauls,rice95}
Along the same line of arguments, 
a magnetic field in the plane should also
split $T_c$, as was pointed out in Ref.~\onlinecite{agterberg98}.

\subsection{Thermal conductivity}

The in-plane thermal conductivity
is isotropic for all order parameter models listed in Table~\ref{basis},
assuming a cylindrical Fermi surface. 
In the clean limit, $T^* \ll T \ll T_c$,
and neglecting logarithmic corrections, 
$\kappa_{i i} \sim T$ for weak scattering and 
$\sim T^3$ for strong scattering.  
In the dirty limit, $T \ll T^* \ll T_c$, the thermal conductivity is linear
in temperature, $\kappa_{i i} \sim T$, and independent of the scattering 
strength.
Unfortunately the samples studied by Suderow {\it et al.}\cite{suderow98}
exhibit a very strong $T_c$ suppression.
The reported resistive transitions for samples \#2 and \#4,
$T_c^\varrho(\#2) \approx 0.81\, {\rm K}$ and
$T_c^\varrho(\#4) \approx 0.58\, {\rm K}$,
occurred significantly above the bulk superconducting
transitions identified by the thermal conductivity,
$T_c^\kappa(\#2) \approx 0.60\, {\rm K}$ and
$T_c^\kappa(\#4) \approx 0.47\, {\rm K}$.
Not only does this suggest that the samples are in the dirty limit
but also that they are considerably inhomogeneous.
Thus the standard scattering $t$-matrix analysis in terms of pointlike 
defects in the dilute limit will most likely fail to give a quantitative 
description. 
Nevertheless, combining the facts of the $T_c$ suppression and that 
the ratios of the residual resistivities and the normal-state thermal 
conductivities are related to the scattering rates,
$\Gamma(\#4) / \Gamma(\#2) \sim \varrho_0(\#4) / \varrho_0(\#2)
 \sim \kappa_N(\#2) / \kappa_N(\#4) \approx 1.25$,\cite{suderow98} 
we find that the normal-state scattering rates are approximately
given by
$\Gamma(\#2)/\pi T_{c0} \approx 0.20$
and
$\Gamma(\#4)/\pi T_{c0} \approx 0.25$
(see Table~\ref{pairbreaking} for the corresponding $T_c$ suppression).

\noindent
\begin{minipage}{86mm}
\noindent
\begin{table}
\caption[]{$T_c$ suppression due to the pair-breaking effects of 
nonmagnetic impurities after Abrikosov and Gorkov.\cite{abrikosov60}
}
\label{pairbreaking}
\begin{tabular}{c|ccccccc}
$\Gamma/\pi T_{c0}$	& 0.0 & $10^{-3}$ & $10^{-2}$ & 0.10 & 0.20 & 0.25
\\ \hline
$T_c(\Gamma) / T_{c0}$		& 1.0 & 0.998 & 0.98 & 0.74 & 0.44 & 0.25
\\
$T_c(\Gamma) \, [K]$		& 1.51 & 1.507& 1.48 & 1.12 & 0.67 & 0.38
\end{tabular}
\end{table}
\end{minipage}

\begin{figure}
\noindent
\begin{minipage}{86mm}
\epsfysize=80mm
\centerline{ \rotate[r]{\epsfbox{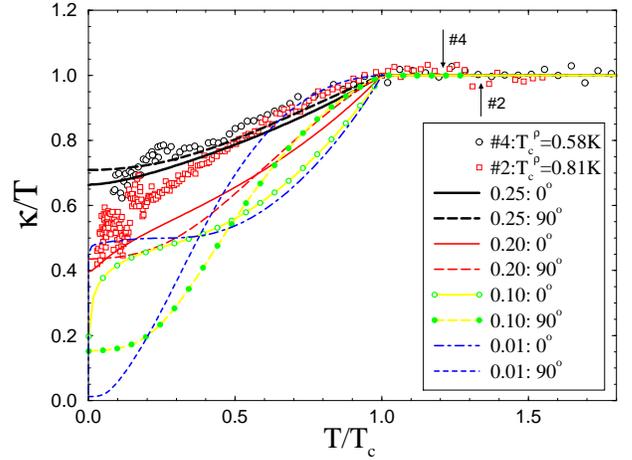}} }
\caption[]{The thermal conductivity normalized at $T_c$
for the pairing state \#1 or \#3.
The scattering phase shifts are
$\delta_0= 0^o, 90^o$, the scattering rates are
$\Gamma/\pi T_{c0} = 0.01, 0.10, 0.20, 0.25$, and
$\mu = 1.5$.
The data are from Ref.~\onlinecite{suderow98}. 
Note that in these dirty samples the resistive 
transition occurs at much higher temperatures (see arrows).
}\label{thermal1}
\end{minipage}
\end{figure}

\begin{figure}
\noindent
\begin{minipage}{86mm}
\epsfysize=80mm
\centerline{ \rotate[r]{\epsfbox{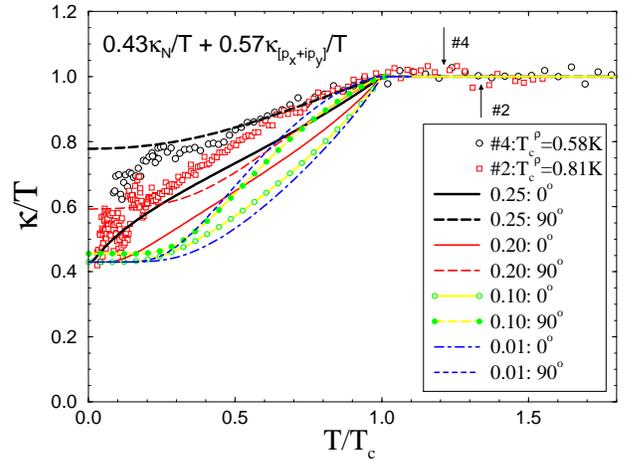}} }
\caption[]{The thermal conductivity normalized at $T_c$
for the multiband model
by Agterberg
based on the $p$-wave state (\#2) and
for the same parameters as in Fig.~\ref{thermal1}.
}\label{thermal2}
\end{minipage}
\end{figure}

In Figs.~\ref{thermal1} and \ref{thermal2} 
we show the best fits of $\kappa_{x x}$ for samples
\#2 and \#4 measured by Suderow et al.\cite{suderow98}. 
Although we cannot obtain a 
quantitatively good fit for any of the pairing
models, we are able to ascribe the large residual value of $\kappa/T$
to impurity scattering (see Fig.~\ref{thermal1}) without having to invoke
a multiband order parameter model (see Fig.~\ref{thermal2}).
A surprising result of these fits is that, generally, we find 
better agreement between theory and experiment for weak impurity
scattering in the Born limit.
Very recently, Tanatar et al.\cite{tanatar00} reported measurements
of $\kappa$ on cleaner crystals ($T_c \approx 1.4\,{\rm K}$) that
are in good quantitative agreement with the gapless states 
\#1 or \#3 and impurity scattering in the unitarity 
limit.\cite{graf_unpublished}

For the predicted pairing states \#1 or \#3, we expect to observe a
fourfold oscillation of the thermal conductivity when a magnetic
field is parallel to the layers and rotated within the layers.
However, the amplitude of the oscillations depends on the scattering
strength. It is appreciable for strong scattering (unitarity limit)
and very small for weak scattering (Born limit).
So far, no oscillations have been observed.\cite{tanatar00,matsuda00}
Certainly the experimental and theoretical situation remains unresolved 
and requires more study.
Indeed such magnetic oscillations have been reported in the cuprate
YBa$_2$Cu$_3$O$_7$,\cite{yu95,aubin97,hirschfeld98,vekhter99} 
and are considered as additional proof in support of the
$d_{x^2-y^2}$ symmetry of the superconducting state.

\subsection{Sound Attenuation}

\begin{figure}
\noindent
\begin{minipage}{86mm}
\epsfysize=85mm
\centerline{ \rotate[r]{\epsfbox{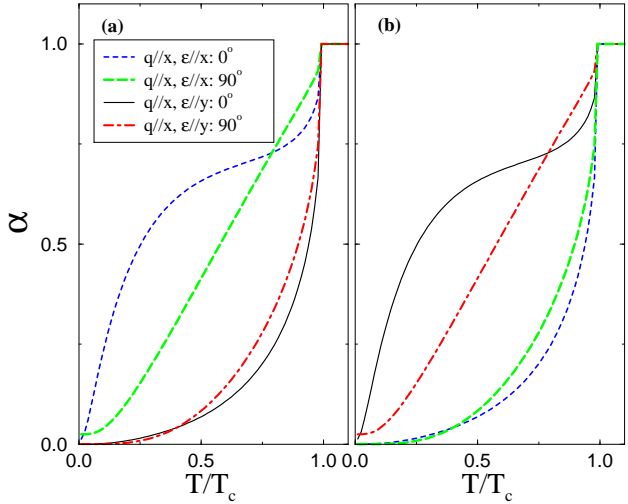}} }
\caption[]{The longitudinal and transverse sound attenuation 
normalized at $T_c$ for the states \#1 or \#3, 
$\Gamma / \pi T_{c0} = 0.01$, $\delta_0=0^o, 90^o$, and
$\mu=1.5$. 
In panel (b) the crystal (or detector) has
been rotated by $\pi/4$ around the $\vc$-axis relative to
the arrangement in panel (a).
}\label{sound}
\end{minipage}
\end{figure}

The longitudinal ($\vq || \vvarepsilon || [100]$)
and transverse  ($\vq || [100]$ and $\varepsilon || [010]$)
sound attenuations are identical for the
pairing state \#2, i.e., 
$\alpha_{xx}(T)/\alpha_{xx}(T_c) = \alpha_{xy}(T)/\alpha_{xy}(T_c)$.
This result also was reported in Ref.~\onlinecite{kee99}.
Whereas for pairing states \#1 and \#3 the longitudinal attenuation with
$\vq || \vvarepsilon || [100]$ is the same as the transverse attenuation
rotated by $\pi/4$ with $\vq || [110]$ and $\vvarepsilon || [\bar{1}10]$.
These relations follow directly from Eq.~(\ref{viscosity})
and are a peculiarity of the 2D Fermi surface and the 2D basis functions  
of the order parameters. Inspecting the momentum dependent
weighting factors in Eq.~(\ref{viscosity}), 
\begin{equation}
\displaystyle
\pi_{xx}^2 = \mbox{$\scriptsize 1\over 4$} \cos^2 2\phi \,, \quad
\pi_{xy}^2 = \mbox{$\scriptsize 1\over 4$} \sin^2 2\phi \,,
\end{equation}
it is clear that by rotating the crystal (or the transducer) by $\pi/4$
around the $\vc$-axis one simply exchanges these functions,
$\pi_{xx}^2 \leftrightarrow \pi_{xy}^2$, and, thus swaps the expressions for
the longitudinal and transverse attenuation. 
Since the integrand ${\cal K}(\vp_f,\epsilon)$ 
for the $p$-wave state (\#2) is independent of $\vp_f$, 
the longitudinal and transverse attenuations are identical (within an overall
scaling factor due to differences in the speed of sound)
for arbitrary temperature and impurity concentration.  These predictions
should be straightforward to check experimentally.
In Fig.~\ref{sound} we show the predicted transverse and longitudinal sound
attenuations for the $d$-wave (\#1) and $f$-wave (\#3)
order parameter models. 
Our results are similar to the ones discussed by Moreno and 
Coleman\cite{moreno96} for the case of the $d_{x^2-y^2}$-wave gap function 
in the high-$T_c$ cuprates.

\section{Conclusions}

We have proposed a spin-fluctuation model based on the measured
spin susceptibility by neutron scattering that leads to nodes
of the gap function on the Fermi surface. We demonstrated that
the measured specific heat  and thermal conductivity are consistent with
a spin-singlet order parameter ($d_{xy}$-wave symmetry belonging
to $B_{2g}$) or a spin-triplet order parameter ($f$-wave symmetry
belonging to $E_u$), though inconsistent with a gapped
spin-triplet state ($p$-wave symmetry belonging to $E_u$).
Based on this analysis we proposed sound attenuation
measurements and thermal conductivity measurements in a magnetic
field to locate the nodes on the Fermi surface, as well as
measurements of the specific heat subjected to a uniaxial 
strain field in the plane in order to split the superconducting transition.
It is clear that more experiments are needed to investigate the
nodal regions on the Fermi surface and the spin structure of
the order parameter.

\acknowledgments
We are indebted to  J.A. Sauls and L. Taillefer for many insightful 
discussions and thank Y. Maeno, M. Sigrist, and D. Agterberg
for discussions.
We thank M. Tanatar and Y. Matsuda for sharing their data prior
to publication.
We acknowledge the Aspen Center for Physics for its hospitality.
This work was supported by the Los Alamos National Laboratory 
under the auspices of the US Department of Energy.


\end{multicols}

\end{document}